\documentclass[
    ,final            
  ]
  {aipproc}

\layoutstyle{6x9}

\begin{document}

\title{A review of forward proton tagging at 420m at the LHC, and relevant results from the Tevatron and HERA}

\author{Brian Cox}{
  address={Department of Physics and Astronomy, The University of Manchester, Oxford Road, Manchester, M139PL, UK}
}

\begin{abstract}
We review the theoretical and experimental motivations behind recent proposals to add forward proton 
tagging detectors to the LHC experiments as a means to search for new physics. 
We also review the current diffractive programs  at the Tevatron and HERA, focusing in particular on
measurements which will have a direct impact on the case for forward detectors at the LHC.    

\end{abstract}

\maketitle


\section{Introduction}

Diffractive physics has provided a rich supply of results from both HERA and the Tevatron. 
As an example, the measurement 
of the diffractive proton structure function from H1 \cite{Adloff:1997sc} is, at the time 
of writing, the collaborations' second highest cited paper. It is fair to say, however, that diffraction has been 
used primarily as a tool for understanding and developing QCD, rather than as an area of study within which 
physics beyond the Standard Model might appear. There has been increasing interest in the past few 
years, however, in the possibility of using diffractive interactions as a search tool for new physics. 
In particular, it has been suggested that the so-called central exclusive production process might provide a 
particularly clean environment to search for, and identify the nature of, new particles at the LHC.
By central exclusive, we refer to the process $pp\rightarrow p \oplus \phi \oplus p$, where 
$\oplus$ denotes the absence of hadronic activity (`gap') between the outgoing protons and the 
decay products of the central system $\phi$. An example would be Standard Model Higgs Boson production, 
where the central system could consist of 2 b-quark jets, and {\it no other activity}. What is meant by 
{\it no other activity} is an important question, and we shall return to it later.   

The process is attractive for two main reasons. Firstly, if the outgoing protons remain intact and scatter
 through small angles, then, to a very good approximation, the central system $\phi$ is predominantly produced in a spin $0$,
 CP even state, therefore allowing a clean determination of the quantum numbers of any observed resonance. Secondly, 
as a result of these quantum number selection rules, coupled with the (in principle) excellent mass resolution on the
 central system achievable if suitable proton detectors are installed, signal to background ratios of greater than unity 
are predicted for Standard Model Higgs production \cite{DeRoeck:2002hk}, and significantly larger for the lightest Higgs 
boson in certain regions of the MSSM parameter space \cite{Kaidalov:2003ys}. Simply stated, the reason for these large 
signal to background ratios is that exclusive $b$ quark production, the primary background in light Higgs searches, is 
heavily suppressed due to the quantum number selection rules. Another attractive feature is the ability to directly probe 
the CP structure of the  Higgs sector by measuring azimuthal asymmetries in the tagged protons (a measurement previously 
proposed only at a future linear collider) \cite{Khoze:2004rc}.

Given the apparent benefits of the central exclusive process at the LHC, two key questions 
naturally arise. Firstly, do we understand diffractive processes well enough to use them as a search tool? This 
applies not only to the calculation of the production rates of new particles, but also to the potential backgrounds, 
most of which also come from diffractive (though not necessarily {\it exclusive}) processes? Secondly, is it possible 
to install leading proton detectors with the appropriate acceptance, and if so, be able to use such detectors at high 
luminosity and integrate them with the existing ATLAS and / or CMS trigger frameworks? In this paper we will review
the current predictions for signal and background rates for a variety of physics scenarios in central exclusive 
production at LHC. We then survey the  
measurements currently being made at the Tevatron and HERA which will have a direct impact on the proposal to install 
forward proton taggers for the detection of the central exclusive process.

\section{Central production at the Tevatron and LHC}

\begin{figure}
\caption{The exclusive production process (a) and central inclusive (or double pomeron) process (figures taken from 
\cite{Khoze:2002py})}
\includegraphics[height=.5\textheight]{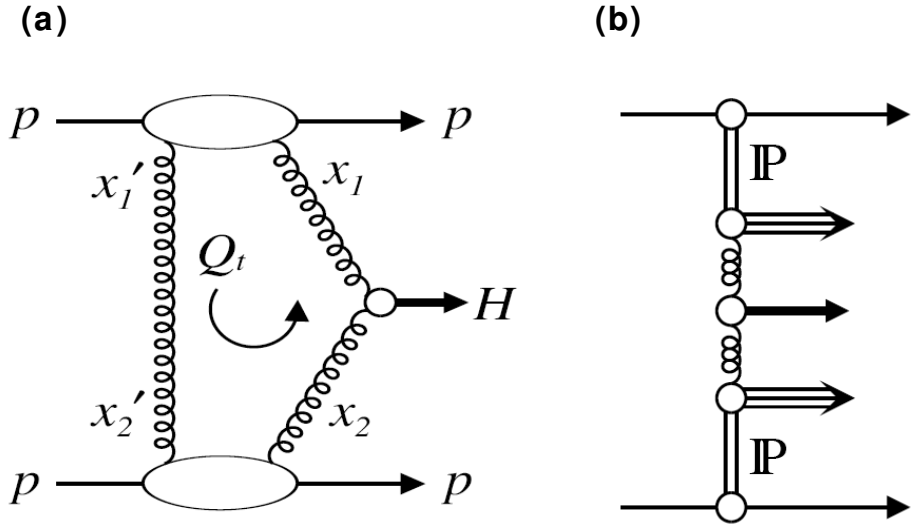}
\label{exinc}
\end{figure}

There has been a healthy debate over the last few years regarding the many and different predictions for the 
production rates of Higgs bosons with two leading protons. Perhaps the clearest review of the competing models can be found in 
\cite{Khoze:2002py}. Here, the authors are careful to differentiate between {\it exclusive} production - the 
process described in the introduction above, and {\it central inelastic} production. Central exclusive production, 
shown schematically in figure \ref{exinc} (a) for the example of the Higgs, refers strictly to the process in which 
the incoming protons emerge into 
the final state intact, 
having lost a small fraction of their energy, and the only other final state particles are the decay products of the 
Higgs Boson. Central inelastic production, shown in figure \ref{exinc} (b), refers to every other process in
which the outgoing protons remain intact - in the language of Regge-inspired models, there are pomeron remnants. 
Two important facts are worth noting. Firstly, central inelastic production is the only process so far unambiguously observed 
at Tevatron energies, and secondly, it is of no use in the search for new physics at the LHC. Having said this, however,
understanding the central inelastic process is essential, because it will be the dominant source of background to the 
exclusive production process. We deal with this issue later. First, we review the signal and background
 predictions for the exclusive process at the LHC.

\subsection{Predictions for exclusive production of Higgs Bosons at the LHC}

 It is the claim of Khoze et al. that the central exclusive process is perturbatively calculable, up to 
the un-integrated off-diagonal parton distributions of the proton (oduPDFs) \cite{Khoze:2001xm}, 
and the so-called gap survival factor
which accounts for the probability that there are no interactions between the spectator partons in the protons, which would 
destroy the protons and the gaps.    
Sudakov factors, which enter via the requirement that there be no radiation off the three exchanged gluons (see figure \ref{exinc} (a)), ensure
 that (at least for scalar Higgs production) the dominant contribution to the integral over $Q_t$ comes from the region in which $Q_t$ is
perturbatively large. 
This is to be contrasted with earlier approaches in which an infra-red cut-off was introduced 'by hand' and tuned to fit the 
total pp cross section \cite{Cudell:1995ki,Levin:1999qu}. The earliest calculation of the process was carried out by 
Bialas and Landshoff \cite{Bialas:1991wj}, and has recently been updated by Boonekamp et al.
\cite{Boonekamp:2003wm,Boonekamp:2004nu}. The early predictions for the exclusive Standard Model Higgs production 
cross section at 14 TeV were
 all extremely large (over 100 fb). If these predictions are correct, then detection of the Higgs using proton tagging will be 
very easy indeed \footnote{It is very likely that such large cross sections are ruled out by the CDF exclusive $\chi_C$ results discussed in the following section}
. The predictions of Khoze et al. are orders of magnitude lower.
Fortunately, the calculations (approximately) factorise into a luminosity function, which contains the physics of the
colour-singlet gluons, and a hard sub-process cross section. It is therefore possible to check the calculations by observing 
the exclusive production of other higher rate processes. We review the first experimental attempts to do so at the 
Tevatron below.    

According to Khoze et al. the cross section prediction for the production of a 120 GeV Standard Model Higgs boson at 14 TeV is 
3 fb \cite{Khoze:2001xm}\footnote{The modified Bialas and Landshoff prediction, taking into account rapidity gap survival and 
other factors, leads to similar predictions \cite{Boonekamp:2004nu}}.
Apart from the estimates of gap survival probability (which are included in this calculation), 
one of the dominant uncertainties in the above calculation comes from
the knowledge of the oduPDFs. Khoze et al. estimate the uncertainty to be a factor of 2, although a recent study by 
L\"onnblad and Sj\"odahl suggests that the uncertainty may be rather larger (but below a factor of 10) \cite{Lonnblad:2003wx}.
We take this prediction as the benchmark result. For 30fb$^{-1}$ of LHC luminosity, therefore, one would expect $\sim 90$
signal events. This is a small number, so the viability of detection depends crucially on the acceptance of 
the proton detectors, the efficiency of the trigger, and the magnitude of the background. De Roeck et al.
 have made a detailed study, including calculations of the $b \bar b$ backgrounds, the $b$-tagging 
efficiency and the acceptance and mass resolution of possible proton tagging detectors at LHC. 
$b$-tagging is necessary because the exclusive production of gluon jets is not suppressed and therefore has an extremely 
large rate which would totally swamp the Higgs signal.
The bottom line is that, for a luminosity 
of 30 fb$^{-1}$, De Roeck et al. expect 11 signal events, with a signal to background ratio of order 1. Details can be found in \cite{DeRoeck:2002hk},
but we make a few remarks here. The very low $b \bar b$ backgrounds are a result, as mentioned in the introduction, of the 
spin selection rules which are a consequence of the colour-singlet configuration of the exchanged gluons (and the small 
transverse momenta of the outgoing protons). These selection rules are not exact: in fact the $b \bar b$ background is 
proportional to $m_b^2 / E_T^2$, where $E_T$ is the transverse energy of the $b$ jets (which will be of order $m_H/2$).
This is a small effect for a 120 GeV Higgs, but as we shall see, can be important for lighter Higgs bosons which might occur
in certain regions of the MSSM parameter space. The selection rules can also be violated by higher order gluon emission. De Roeck et al. 
estimate the contributions from NLO and NNLO diagrams, and find them negligible. This result can (very crudely) be
pictured as the statement that soft gluons do not flip quark helicities. There may be an issue here, however, as to what 
one means {\it experimentally} by a soft gluon. If, for example, a gluon emitted from an out-going $b$ quark with a relative 
$p_T \sim 4$ GeV is sufficient to violate the selection rules, and yet cannot be resolved experimentally, then what is the
resulting change in the background estimates? This issue has yet to be addressed in detail.

The mass acceptance and resolution of the forward proton detectors is also a crucial issue, which depends on many factors 
including the LHC beam optics, the 
distance of the detectors from the interaction point, the closeness of the active region of the detectors to the beams, 
and the accurate knowledge of the relative positions of the detectors to the beams. De Roeck et al. consider the case in 
which detectors are placed at 420m from the interaction point. This position is simply the distance 
from the interaction point, with 
standard LHC beam optics, that protons which lose transverse momentum $m_H/2 \sim 60$ GeV emerge at least $10 \sigma$ from the beam. 
This large distance raises a serious issue. Without modification of the level 1 
trigger systems of ATLAS and CMS, the light travel time from 420m detectors is very close to, and possibly larger than, the 
time required for a level 1 trigger decision. 
This means that a trigger strategy based on the central detectors alone may be required, at least until
the proton tagger information becomes available at level 2. For dijets of such low transverse momentum ($\sim 60$ GeV), 
this is certainly a 
challenge. Both De Roeck et al. and Boonekamp et al. \cite{Boonekamp:2004nu} consider some basic ideas based on the 
central system topology, but it is fair to say that much work still needs to be done in this area. 

The resolution of the detectors is a crucial number. The signal to background 
$S/B \propto \Gamma(H\rightarrow gg)/\Delta M \propto G_F M^3_H / \Delta M$, where $\Delta M$ is the mass window within which 
the search is performed. This is easily seen: a search using this technique is simply a counting experiment within a mass window,
 and since the tagger resolution will always be greater than the Standard Model Higgs width, the worse the resolution the 
more continuum background will enter \footnote{It is worth noting that in certain MSSM scenarios, the Higgs width can exceed the resolution of the tagging detectors, 
raising the possibility of a direct measurement of the width}. 
         
There is also the question of contamination from the central inelastic process, shown in figure \ref{exinc} (b). As
mentioned above, this is the only process so far observed at the Tevatron \cite{Affolder:2000hd}. There are no selection rules suppressing $b \bar b$ 
production in this process, and it is therefore potentially a very large source of background. We consider this in the
following section. 
          
Finally, we briefly review two other scenarios in which forward proton tagging may be of significant interest at the LHC. 
Firstly, the `intense coupling' regime of the MSSM. This is a region of MSSM parameter space in which the couplings of 
the Higgs to the electroweak gauge bosons are strongly suppressed, making discovery challenging at the LHC by conventional 
means. The rates for central exclusive production of the two scalar MSSM Higgs bosons can be enhanced by an order of magnitude
in these models, however, leading to predicted signal to background ratios in excess of 20 for masses around 
130 GeV\cite{Kaidalov:2003ys}. This region of 
parameter space can also be problematic in conventional search channels because the masses of the three neutral Higgs Bosons
are close to each other. Central production can help disentangle the Higgs bosons because, due to the spin/parity selection rules,
production of the pseudo-scalar (A) Higgs is heavily suppressed. This 
means that the `double tagged' sample will be almost pure scalar. 

As a second example, Higgs sectors with explicit CP-violation \cite{APLB} are also an area in which central production may prove 
extremely attractive. One such model, known as the CPX scenario \cite{Carena:2000ks}, has been shown to lead to very light (less 
than 60 GeV) Higgs bosons which would have evaded detection at LEP, and may well evade detection at the Tevatron or LHC, again 
primarily due to the suppression of the coupling to the electroweak bosons of the lightest Higgs (which may be 
predominantly pseudo-scalar, since the mass eigenstates are not weak eigenstates in this case)
\cite{Carena:2002bb}. The central production cross sections for the lightest CPX Higgs are relatively large at low masses \cite{Cox:2003xp}, 
although the relaxing of the $b\bar b$ background suppression with mass (see above) probably means that the $b \bar b$ decay mode
will not be a possible detection channel. The $\tau \tau$ mode may be possible, 
however, since the only background comes from QED production of $\tau$ 
pairs, which can be suppressed by demanding that the $p_T$ of the tagged protons be greater than a few hundred MeV \cite{Khoze:2004rc}. 
It was also noted in \cite{Khoze:2004rc} that explicit CP violation in the Higgs sector will show itself directly as a (potentially sizeable) 
 asymmetry in the azimuthal distribution 
of the tagged protons. This measurement is probably unique at the LHC, although little detailed  phenomenological 
work has been done so far. First indications suggest that this may be a high-luminosity ($300 \rm{fb}^{-1}$) measurement, although 
models other than CPX have not been considered, and may possibly lead to higher signal to background ratios in the $b \bar b$ channel.  
               
\subsection{Predictions for the central inelastic process}    

Central inelastic production is usually referred to as `double pomeron exchange', and can be modeled using a Regge
 - inspired 
 picture involving a pomeron flux term describing the 'emission' of a pomeron from the proton, and a pomeron structure 
function. The parameters of the flux term and the structure function are extracted from  
diffractive deep inelastic scattering data by the H1 Collaboration at HERA \cite{Adloff:1997sc} using a Regge 
factorisation ansatz\footnote{H1 also require the addition of a reggeon term to fit the data at larger proton 
fractional momentum losses}. This approach has been shown not only to describe a wide range of HERA data, but also 
the double diffractive dijet production data from the CDF Collaboration\cite{Affolder:2000hd,Appleby:2001xk}, if a `gap
 survival factor' is included to account for the fact that multi-parton interactions in proton-antiproton collisions 
will reduce the observed rate of double pomeron collisions. This factor can be extracted from data, by scaling the
absolute predictions derived using the H1 diffractive structure functions \cite{Cox:2001uq}, 
or calculated using phenomenological 
approaches of varying degrees of sophistication, but in general based on eikonal methods and total cross section 
measurements \cite{Dokshitzer:1991he,Bjorken:1991ft,Fletcher:1993ij,Lungov:1995iq,Gotsman:1998mm,Gotsman:1999xq}. It 
is the latter calculational approach that Khoze et al. use to estimate the gap survival probability in their central exclusive predictions.
At least at Tevatron energies, the two
approaches lead to similar results; the Regge-inspired predictions, normalised to the H1 diffractive DIS data, overshoot the 
measured double pomeron exchange data from CDF by a factor of approximately 10 \cite{Appleby:2001xk,Cox:2001uq}, and the 
predictions from the eikonal models are of a gap survival factor of approximately 0.1 \cite{Khoze:2000cy,Kaidalov:2001iz,Kaidalov:2003gy}. 
This gives at least some confidence in the calculations of gap survival probability which are an important ingredient in predicting 
the central exclusive Higgs production rates at the LHC. 

As mentioned above, understanding the central inelastic process is of prime importance. The cross section for the production 
of $b \bar b$ jets in central inelastic production is orders of magnitude larger than the exclusive process. 
The key to suppressing this background is to reject events in which there are pomeron remnants, either directly using the 
central detectors, or by 
requiring that the invariant mass of the dijet system is equal to the invariant mass of the central system as measured in the 
proton taggers. It is the latter solution that was used by De Roeck et al. \cite{DeRoeck:2002hk}. This approach depends on 
defining which final state particles are inside the jets, and which are not, and is clearly extremely sensitive to the 
nature of the jet algorithm used. The CDF collaboration are currently using a similar method to search for evidence of the
exclusive process. They define the observable $R_{jj}$ as being the fraction of the mass of the central system contained within 
the two highest $E_T$ jet cones of radius $R_{cone}=0.7$. Exclusive dijet production would be expected to appear 
in the high $R_{jj}$ tail of this distribution. CDF measure the cross section for $R_{jj} > 0.8$, and find it to be consistent 
with the Khoze et al. predictions, although there are at present large systematic errors and theoretical uncertainties \cite{lefevre}.

An alternative approach is to carry out a more sophisticated analysis on the central system itself - in essence using a subjet
analysis to identify the presence of pomeron remnants. It is our opinion that such an approach will also 
be necessary to address 
the issue of moderate $p_T$ gluon emission reducing the effectiveness of the selection rules, as mentioned above. 
This work is ongoing at the time of writing, and we intend to return to it in a future publication \cite{coxpilkington}.     

\section{Recent results from the Tevatron and HERA}
Clearly, the understanding of central dijet production is crucial for central Higgs production searches at the LHC. In 
terms of searching for evidence of exclusive production at the Tevatron, however, it is not ideal, given the difficulties
in defining which particles are inside the jets. The CDF collaboration have therefore also begun a search 
for the exclusive production of lighter particles with the same quantum numbers as the Higgs. The lightest detectable particle 
(and therefore that with the highest cross section) is the $\chi_c$, which is detected via its decay to $J/\psi$ $\gamma$, and 
the subsequent leptonic decay of the $J/ \psi$. Although this process appears rather simpler than the search for exclusive 
dijets, one still has to face the issue of defining {\it experimentally} what is meant by exclusive: it is feasible that 
soft radiation may go undetected in the calorimeter, and it is extremely difficult to put a figure on what is meant by `soft'. 
For this reason, the CDF Collaboration present their results as an upper limit on the exclusive $\chi_c$ 
production cross section \cite{lefevre}, assuming that all the $\chi_c$ candidates which appear to have no 
other activity in the calorimeter are 
indeed truly exclusive. The result is consistent with the estimates of Khoze et al. although due to the low mass of the 
$\chi_c$, the theoretical calculations are subject to large uncertainties \cite{Khoze:2004yb}. Exclusive diphoton production 
($p\bar p \rightarrow p \gamma \gamma \bar p$ is also a promising channel which could be used as a `standard candle' to 
check the calculations \cite{Khoze:2004ak}. The central exclusive predictions are in the 10s of fb for photons 
of $E_T \sim 7$ GeV, whilst the central inelastic cross section is $\sim 100$fb \cite{Cox:2001uq}. This process should 
therefore provide a clean testing ground not only for the theory, but also for the experimental techniques used to 
separtate the exclusive and inelastic processes.           

Another key area in which the Tevatron and HERA can make valuable contributions is in the testing of the models 
of gap survival probability. We have already considered the 
double diffractive dijet measurements from CDF, \cite{Affolder:2000hd,Appleby:2001xk}, and concluded that in this case a 
consistent picture emerges between theory and experiment. Gap survival is also an issue in diffractive photoproduction
at HERA, in which the resolved photon can behave as a hadronic object and therefore contain spectator partons which can interact 
with and destroy the diffracted proton just as at the Tevatron. The H1 and ZEUS Collaborations have measured the diffractive 
dijet cross section in photoproduction as a function of $x_\gamma$, the fractional longitudinal momentum of the photon
 which participates in the production of the two highest $E_T$ jets \cite{h1diffdijets1,h1diffdijets2,zeusdiffdijets}. Without taking account of gap
 survival, the naive expectation would be that the predictions should match the data at high $x_\gamma$, where the photon 
couples directly into the jet production process, whereas at low $x_\gamma$, where spectator partons are present, the 
predictions should overshoot the data. A recent phenomenological study, which took account of NLO effects, found that 
at low $x_\gamma$ a suppression factor $S=0.34$ was required to fit the data \cite{Klasen:2004qr} \footnote{This result has recently 
been confirmed by the H1 and ZEUS collaborations, with slightly varying figures for the suppression factor \cite{h1diffdijets2,zeusdiffdijets}}. 
This is in agreement with the calculations of \cite{Kaidalov:2003xf}.

To summarise, all the diffractive results from HERA and the Tevatron are at the time of writing consistent with the 
expectations from the theory of exclusive and inclusive  production and gap survival probability. However, this is not 
to say that the data as it now stands should give us blind confidence in the predictions at the LHC. The exclusive process has 
certainly not been unambiguously observed, although with increased luminosity at the Tevatron Run II it certainly should be
seen within the next few years if the calculations are correct.      

\section{Summary}

The installation of proton tagging detectors in the 420m region around ATLAS and / or CMS would certainly add unique capabilities 
to the existing LHC experimental programme. If the current calculations of central exclusive production rates survive the experimental 
tests at the Tevatron, then there is a very real chance that new particle production could be observed in this channel. For the 
Standard Model Higgs, this would amount to a direct determination of its quantum numbers, with an integrated luminosity of order 
30 fb$^{-1}$. For certain MSSM scenarios, the tagged proton channel may be the discovery channel. At higher luminosities, proton tagging 
may provide direct evidence of CP violation within the Higgs sector. There is also a potentially rich, more exotic physics menu which we 
have not discussed, including gluino and squark production, gluinoballs, and indeed any object which has $0^{++}$ or $2^{++}$ quantum 
numbers and couples strongly to gluons \cite{Khoze:2001xm}. Given the relatively low cost of such a project, and the potentially unique 
access to new physics, we believe the installation of 420m proton detectors at LHC should be given careful consideration.         
\begin{theacknowledgments}
We would like to thank Valery Khoze and Jeff Forshaw for many useful conversations and suggestions. 
We would also like to thank the UK Particle Physics and Astronomy Research Council for funding this work. 
\end{theacknowledgments}

\end{document}